\documentclass[cameraready]{Interspeech}
\title{Balancing ASR and diarization in end-to-end LLMs for multi-talker speech recognition}
\author{Naijun}{Zheng}
\author{Yuke}{Lin}
\author{Sanli}{Tian}
\author{Mengtian}{Li}
\author{Zhiwei}{Lin}
\author{Longshuai}{Xiao}
\author{Dandan}{Tu}
\address{
    $^1$ Huawei Technologies, China
}

\email{njzheng.cuhk@gmail.com}


\usepackage{comment}
\usepackage{CJK}
\usepackage{arydshln}

\begin{document}
% \begin{CJK}{UTF8}{gbsn}

\maketitle

\begin{abstract}  
    Multi-talker speech recognition is often addressed by combining automatic speech recognition (ASR) and speaker diarization in a pipeline system.
    Recently, LLM-based approaches have shown promise by jointly modeling semantic and speaker information, but they typically require large-scale multi-talker corpora that are costly to annotate.
    In this paper, we investigate how to efficiently train an LLM-based system with limited real-recorded data while maintaining high accuracy in speaker attribution. % under overlaps.
    % Multi-talker speech recognition is commonly addressed as a combination of automatic speech recognition (ASR) and speaker diarization, which can be addressed by aligning the diarization results and transcript hypothesis in a pipeline system.
    % Recently, large language model(LLM)-based approaches have shown promise by jointly modeling semantic and speaker information. 
    % However, training such systems usually requires large-scale multi-talker corpora, which are costly to annotate.
    % While it follows a data scaling law, where thousands hours corpus are required to teach the LLM to understand both acoustic information and semantic information, especially how to deal with the overlaps and speaker turns in the dialogues.
    % Convention approaches utilize . %when meeting overlaps and speaker turns.
    % How to efficiently utilize these features in an end-to-end mode is a key issue. % to beat the cascaded systems.
    % In this paper, we investigate how to efficiently train an LLM-based system with small-scale real-recorded data, while maintaining high accuracy in identifying speakers within turns and overlaps.
    We propose several strategies:
    (1) a dual-encoder architecture to extract semantic and speaker features,
    (2) a feature interleaving format to merge these features as the inputs to the LLM,
    (3) a length-aware speaker ID loss to enhance diarization capability, and 
    (4) an adaptive threshold strategy for ASR loss computation to mitigate hallucinations caused by speech overlaps.
    % and apply an innovative feature interleaving format to merge them as the input.
    % Meanwhile, a length-aware speaker ID loss is proposed, which improves the diarization capability.
    % To alleviate the hallucination caused by speech overlaps, an adaptive threshold strategy is applied on ASR loss computation by masking the tokens of high loss values.
    % It can make the model get rid of high token loss within the overlaps and pay more attention to the easy recognized ones. 
    These strategies balance training between ASR and diarization tasks. Our system outperforms open-source baseline approaches, achieving relative improvements of 18\% on the AliMeeting corpus and 24\% on the Aishell4 corpus.
    % our systems balances training between ASR and diarization tasks, outperforming open-sourced pipeline approaches with relative improvements of 18\% on AliMeeting and 24\% on Aishell4 datasets.
    
\end{abstract}

\section{Introduction}

Multi-talker speech recognition\cite{7122291,9072433,9383556} has been widely applied in scenarios such as meeting recordings, subtitle transcription, and dialogue understanding, aiming to solve the problem of "who spoke what." 
Since the task can be split into both automatic speech recognition (ASR) and speaker identification tasks, it can be solved with the pipeline systems composed of ASR modules and diarization modules \cite{Kanda2021TranscribetoDiarizeNS, SCDiar, METACATwang}. These systems are relatively lightweight in parameters and deliver robust performance.
The final speaker-attributed transcripts can be obtained from the speaker diarization results and transcripts based on the aligned time-stamps.
However, since the modules are trained independently, speaker identification is not directly integrated with the corresponding semantic content of the conversation, especially when meeting overlaps \cite{10037902}. This limitation has motivated the use of large language models (LLMs), which can jointly capture speaker roles and dialogue contextual information simultaneously within the model.

% Multi-talker speech recognition task has been widely applied in meeting recording, subtitle transcriptions and dialogue understanding scenarios, which aims to solving the problem "who spoke what".
% Since it combines both ASR and speaker identification tasks, several modules can be simply cascaded to compose a pipeline systems like \cite{Kanda2021TranscribetoDiarizeNS, SCDiar, METACATwang}, which have fewer parameters and robust performance. 
% The speaker id results and transcripts of two tasks are  then aligned on the time-stamps to get the final results.
% % Compared with the recent LLM-based ASR systems, 
% However, due to the dependent-trained modules, the speaker role identification does not considered with the corresponding semantic contents  in the conversations. 
% This motivation push the emergence of using the large language models (LLMs) to dig the potential values within dialogues to understand both the speaker roles and their contexts. 

Early work such as DiarizationLM \cite{wang24h_interspeech} leverages LLMs to correct speaker IDs given multi-talker transcripts, using contextual cues to detect speaker turns. 
Some approaches incorporate existing diarization modules \cite{Sortformer, lin2025diarizationawaremultispeakerautomaticspeech} to  supply speaker-activity information that guides LLM generation. 
More recently, several end-to-end approaches \cite{shi2025trainshortinferlong, huo2026tagspeechendtoendmultispeakerasr, ai2026mosstranscribediarizetechnical} have been proposed to directly infer speaker identities and their corresponding content. % without relying on external diarization predictions. 
SpeakerLM \cite{yin2026speakerlmendtoendversatilespeaker}, trained on over 7,000 hours of real-meeting data, demonstrates excellent performance compared to cascaded pipelines.
TagSpeech \cite{huo2026tagspeechendtoendmultispeakerasr} and Vibevoice-ASR \cite{peng2026vibevoiceasrtechnicalreport} adopt dual-encoder architectures to provide both semantic information and acoustic information for speech recognition and speaker identification.

Nevertheless, training LLMs to robustly model speaker IDs in natural conversations remains challenging, especially in the presence of interruptions, backchannels, and overlapped speech. 
These phenomena are common in multi-talker scenarios but impose high demands on system robustness, which demands large-scale annotated data for training.

% SpeakerLM \cite{yin2026speakerlmendtoendversatilespeaker} trained with more than 7,000 hours real-meeting data, shows excellent performance compared to cascaded pipelines.
% % End-to-end model JEDIS-LLM  \cite{shi2025trainshortinferlong} tagspeech\cite{huo2026tagspeechendtoendmultispeakerasr} with speaker cache.
% Moss-Diarize\cite{ai2026mosstranscribediarizetechnical} and Vibevoice-ASR \cite{peng2026vibevoiceasrtechnicalreport} applied dual-encoders to prove accurate speaker information for LLM.
% However, for such multi-task training, teaching  LLM to learn the speaker roles is not a cheap thing, especially when meeting interruption, backchannel and overlaps.
% This speaker turns are common but propose a high standard for systems which pron to lead hallucination and requires huge amount of data to make the system robust.

In this paper, we propose an efficient approach to balance ASR and diarization tasks through both model structure and loss function design, without relying on large-scale meeting corpus. 
Our contributions are summarized as follows:
% (1) We investigate several feature merge strategies for the outputs of the dual encoders and find that an interleaved format best aligns speaker and speech features, effectively balancing speaker identification and speech recognition tasks.
(1) We explore multiple strategies for merging the outputs of dual encoders and find that an interleaved representation provides the most effective alignment between speaker-related and speech-related features. %, leading to a better balance between speaker identification and speech recognition.
(2) We introduce a segment-aware speaker ID loss function to improve identification accuracy, which better aligns with diarization metrics compared to using ASR loss alone.
(3) %To address overlapped speech in multi-talker scenarios, 
We analyze the ASR loss behavior and observe that the tokens within the overlaps are of high-loss and often trigger hallucinations and repeated outputs. We therefore propose an adaptive loss mask that suppresses high-loss tokens during training, significantly reducing hallucinations.
% Additionally, we apply re-alignment labeling and other training refinements to further enhance performance.
With these strategies, our system achieves significant improvements over strong baselines.
%while avoiding reliance on large-scale real-recorded training data.

% In this paper, we find a more efficient way to balance the two tasks, in term of both model structures and loss function design, instead of collecting real-record data in training.
% Our contributions are listed as follows:
% (1) Based on a dual-encoder structure, several feature merging approaches are investigated carefully, where we find that the interleaved formate can most benefit the alignment between the speaker and speech features well balance the speaker identification task and the speech recognition task.
% (2) Regarding to the loss function, only using ASR loss functions is not enough. We propose a segment-aware speaker id loss function to improve the identification accuracy, which better cope with diarization metrics.
% (3) To tackle the overlapped speech in multi-talker scenarios, we deeply investigate the ASR loss during training and observe that the high-loss token recognition maybe the root cause of the LLM hallucination to generate repeated answer. 
% Thus, an adaptive loss mask is added on the ASR loss during training, which greatly reduce the hallucinate. 
% Other tricks, likes re-alignment labeling, are also applied to benefit the training.
% With these strategies, our system can achieve XX the baselines without using XX real data.

\section{Proposed Method}

\begin{figure}[htp]
    \centering
    \includegraphics[width=1.0\linewidth]{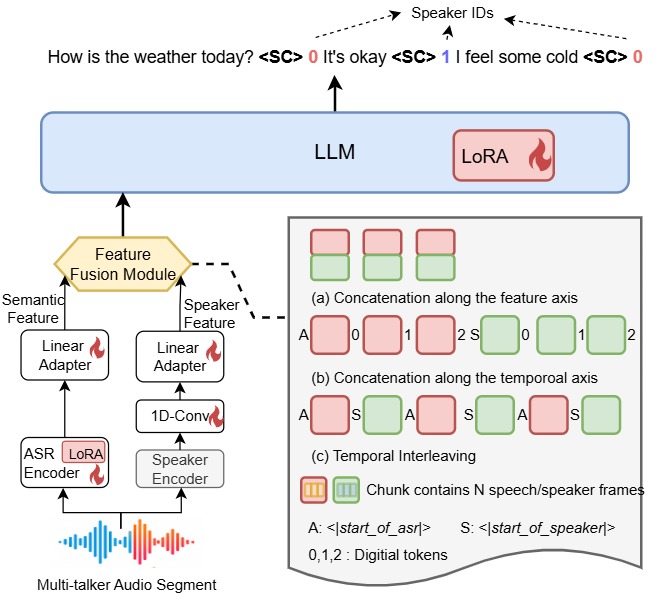}
    \caption{Dual-encoder architecture with different feature fusion strategies.}
    \label{fig:system}
\end{figure}

\subsection{Dual-encoder structure and feature fusion strategies}
\label{ssec:mergy}

Figure \ref{fig:system} illustrates the overall architecture, which consists of two encoders: an ASR encoder and a speaker-feature encoder. 
Given an audio segment containing multiple speakers, the dual encoders extract semantic features and speaker features, which are then combined in various formats before being fed into the LLM backend.

% Figure \ref{fig:system} shows the whole diagram of the system which contains two encoders, one ASR encoder and one speaker feature encoder.
% A audio segment recorded from several speakers goes through the dual encoders to obtain the semantic feature and speaker feature.
% The features can be combined in different format before input to the the LLM backend.

We adopt SenseVoice-small \cite{FunAudioLLM} as the ASR encoder, which is a CTC-based module. 
Hidden states from both its final layer and an intermediate layer are extracted and concatenated along the feature dimension,
which retain both semantic cues and acoustic cues.
% The outputs from shallow-layer retain acoustic cues, enabling the ASR encoder to contribute to both speech recognition and speaker-related tasks. 
Then a lightweight adapter composed of two linear layers with GELU activation is applied for dimension projection to get the semantic feature sequences.

% We apply the SenseVoice-small \cite{FunAudioLLM} as our ASR encoder, which is a CTC-based ASR module.
% Two hidden state outputs from its last layer and middle layer are extracted respectively and concatenated along the feature axis.
% Since the feature from the shallow layer can maintain some acoustic information, the semantic features from the ASR encoder can support both tasks to some extent.
% The linear adapter is composed of two linear layers with the GELU function.

For speaker representation, we employ the Campplus module \cite{wang23ha_interspeech}. We modify the original architecture by inserting a chunking operation before the pooling layer: instead of computing the mean vector and standard deviation vector over the entire sequence, statistics are computed within each chunk. 
The resulting chunk-wised embeddings are then repeated to match the input sequence length.
To balance robustness and temporal resolution, we use three chunk sizes (400 ms, 200 ms, and 100 ms). 
Longer chunks yield more stable speaker embeddings, while shorter chunks provide finer time granularity. 
The embeddings from all chunk sizes are concatenated along the feature axis and passed through projection layers. 
A 1D convolution layer is used to downsample the speaker feature sequence  to match the length of the semantic feature sequence. We freeze the parameters of the Speaker encoder during training.
% As for the speaker encoder, we apply the Campplus module \cite{wang23ha_interspeech} to extract speaker features.
% We modify the original structure by adding a chunk function before the pooling layer, where the standard deviation and mean vectors are computed on the chunked segments rather than the whole sequence.  Then the obtained embeddings are repeated to have the same lengths as input.
% We use three different chunk sizes (400 ms, 200 ms, and 100 ms) before pooling, since the longer speaker segments can obtain more robust speaker embeddings but lower time-resolution.  Then we concatenate all embedding sequences along the feature axis as the speaker features followed by the projection linear layers.
% We also use an 1D-convolution layer to downsample the speaker features to match the same length as the semantic features.
Formally, the feature extraction process is:
\begin{align}
    X_{asr} &=\text{ASRAdapter}( \text{Cat}(\text{ASREncoder}(X))) \\
    X_{spk} &=\text{SpkAdapter}( Conv\text{1D}(\text{Cat}(\text{SpkEncoder}(X)))
\end{align}
Both adapters consist of two linear layers followed by layer normalization.
% Both adapters contain two linear layers and a layer normalization layer.

After obtaining the semantic sequence and speaker‑feature sequence for each sample, the key question is how to merge them into a unified representation as the input for the LLM backend. We explore several fusion strategies, illustrated in Figure \ref{fig:system}.
\begin{itemize}
    \item \textbf{Semantic Feature Only}:  
    Only the semantic feature sequences from the ASR encoder are used as input, serving as a baseline without explicit speaker information.
    \item \textbf{Feature-wise Concatenation}:  
    The semantic and speaker feature sequences are concatenated along the feature dimension and then projected to the target dimensionality.
    \item \textbf{Time-wise Concatenation}:  
    Following TagSpeech~\cite{huo2026tagspeechendtoendmultispeakerasr}, we concatenate the speech and speaker feature sequences along the temporal axis. The same digital tokens are inserted every 20 frames (1.2\,s) into both sequences, and each sequence is prefixed with a special token \texttt{<semantic\_start>} or \texttt{<speaker\_start>}.
    \item \textbf{Temporal Interleaving}:  
    The two feature sequences are interleaved along the temporal axis for each $N$ frames, where $N$ is set to 20 denoting 1.2\,s.
    Each chunks is prefixed with \texttt{<semantic\_start>} or \texttt{<speaker\_start>}. 
    Unlike time-wise concatenation, no digital tokens are used, and the alignment relies solely on positional encoding between adjacent semantic and speaker chunks.
\end{itemize}

% % After obtain the the semantic and speaker feature sequences, a key issue is how to combine them as the input to the LLM. We proposed several merge strategies, as shown in Figure \ref{fig:system}.
% \begin{itemize}
%     \item \textbf{Semantic Feature Only}: We only use single ASR encoder its semantic features as input.
%     \item \textbf{Feature‑wise Concatenation}: We concatenate the semantic and speaker feature sequences along the feature axis, and project the dimensionality using a linear projector.
%     \item \textbf{Time‑wise Concatenation}:   Following Tagspeech\cite{}, we concatenate the speaker and speech feature sequences along the temporal axis, and  insert the same digital tokens embeddings in both speech features and speaker features for each 20 frames (1.2 seconds). An special token \textit{Speech-start} and \textit{Speaker-start} are prefixed for two feature sequences.
%     \item \textbf{Temporal Interleaving}: We concatenated the feature blocks in an interleave formate along the temporal axis. Each block contains 20 frames (1.2 seconds) chunked from two feature sequences. The blocks are prefixed with a special token \textit{Speech-start} or \textit{Speaker-start}. Different from time-wise concatenation, no digital anchor is used, and the adjacent semantic block and speaker blocks are aligned with only position encoding. 
% \end{itemize}
% % Different from the feature-wise concatenation, no dimension projection are applied on semantic and speaker information to undermine their completeness but with longer input.

\subsection{Training process and loss functions}

We adopt a multi-stage training scheme to progressively align the dual encoders with the LLM.
% We design a multiple-stage training scheme to align the encoders to the LLM.

The system is first trained on the ASR task using only semantic features. In this stage, the adapter parameters are trainable, and Low-Rank Adaptation (LoRA)~\cite{hu2022lora} is applied to the LLM.

% In the first stage, we trained the system on the ASR task using only semantic features, where adapter parameters are trainable and the Low-Rank Adaptation (LoRA)  \cite{hu2022lora} is applied on LLM. 
%WenetSpeech\cite{} is used as training data.

In the second stage, speaker features are introduced, and the system is trained jointly on ASR and diarization tasks using a two-speaker conversation corpus. The fusion strategies described in Section~\ref{ssec:mergy} are applied. 
Since speaker change detection can be regarded as an auxiliary subtask of diarization~\cite{7953097, SCDiar}, 
we introduce the special token \texttt{<SC>} into the training labels to explicitly mark speaker transitions. 
This design treats \texttt{<SC>} as a prompt for the model to infer the subsequent speaker identity, 
thus bridging the simpler change-detection task with the more challenging speaker identification task.
% Since speaker change detection is generally simpler than full speaker identification~\cite{7953097, SCDiar}, we insert the special token \texttt{<SC>} into the labels to denote speaker changes. 
The speaker ID is placed immediately after \texttt{<SC>} to indicate the speaker of the preceding recognized segment. Thus, the labels for multi-talker ASR take the form:
% In the second stage, we added the speaker features to the input and train the system on both tasks using a two-speaker conversation corpus.
% In this stage, the merge strategies in Section\ref{ssec:mergy} are applied.
% % We as training data.  from  audio segments.
% Since the speaker change detection can be regard as a simpler task compared to the speaker identification task\cite{}, we add \textless SC\textgreater  in labels to denote the speaker changes in the dialogues.
% Then, we put the speaker ID after \textless SC\textgreater    to denote the speaker of the former recognized context.
% % Thus, we put the spkid after the \textless SC\textgreater to represent the speaker in the former segment.
% Finally, the labels for multi-talker ASR become as: 
\begin{equation} 
    \{ \text{text}_1 \ \texttt{<SC>} \ \text{spk}_1 \} \quad \{ \text{text}_2 \ \texttt{<SC>} \ \text{spk}_2 \}, 
    \label{eq:label} 
\end{equation}

In the third stage, we simulate long conversations by concatenating several segments during training, where each sample contains up to eight speakers. Virtual \texttt{<SC>} tokens are also inserted within long pauses of the same speaker, while keeping the speaker ID unchanged. 
This design prevents false alarm errors in speaker change detection from directly causing errors in speaker ID assignment, which makes the model learn that not every \texttt{<SC>} indicates a new speaker. 
% As a result, false alarms in change detection do not mislead the system into wrongly switching speaker identities.
% This prevents false alarms in speaker change detection from propagating errors into speaker ID inference. 
Moreover, we propose a segment-aware speaker identification loss that better aligns with diarization metrics such as concatenated minimum-permutation character error rate (cpCER) \cite{Watanabe2020CHiME6CT}. As shown in Eq.~\eqref{eq:spkloss}, segment lengths are used as weights when computing the cross-entropy loss for speaker IDs:
% In the third stage, we simulate long samples by concatenating several segments during training, where each sample contains at most eight speakers.
% We also add virtual \textless SC\textgreater within the same speaker segments when meeting long pauses, and the following speaker ID is kept the same.
% % to avoid the errors in \textless SC\textgreater cause the error in speaker ID inference,  since the insert error or miss error in SC will cause the speaker id inference errors,
% In this way,  the false alarm errors in speaker change detection will not mislead the speaker ID identification. 
% % but perform as the prompt to trigger the ID inference.
% Moreover, we propose a new loss function for speaker identification, which can better cope with diarization metrics, such as Concatenated Minimum-Permutation Word Error Rate
% (cpWER). 
% As shown in Eq\eqref{eq:spkloss}, the lengths of speaker segments play as weights when compute the cross entropy for speaker IDs.
\begin{equation}
    L_{spk} = \frac{1}{\sum_{i=0}^{N-1}L_i}\sum_{i=0}^{N-1}L_i*\text{CE}(spk_i),
    \label{eq:spkloss}
\end{equation}
where $N$ is the number of speaker segments in a sample, $L_i$ is the token length of the $i$-th segment, and $\text{CE}(\text{spk}_i)$ is the cross-entropy loss for its speaker ID. 
By assigning larger weights to longer segments, the training objective emphasizes accurate speaker identification in extended turns.
% By weighting longer segments more heavily, the model learns to prioritize accurate identification in extended speaker turns. 
The overall training objective is the sum of the ASR loss and the speaker ID loss.
% where $N$ is the number of speaker segments in a sample,  $L_i$ is the token number of the $i$-th segment and $\text{CE}(spk_i)$ is the cross entropy loss for its speaker ID.
% With considering the segment lengths, the model can pay more attention on long segments.
% Finally, the ASR loss and speaker ID loss are summed together for training.

In the final training stage, the model is fine-tuned on real in-domain meeting corpora, with LoRA applied to both the ASR encoder and the LLM. 
Unlike the simulated training data, real meetings often involve a high degree of overlapped speech. Such overlaps not only increase the error rate in ASR but also tend to induce repetition hallucinations in the LLM during inference. 
% In the following subsection, we investigate the underlying causes of these hallucinations and how to mitigate them.
In the following subsection, we investigate the causes of hallucinations and discuss strategies to mitigate their impact.
% In the last stage, we use the real-recorded in-domain meeting corpus to finetune the adapters and applied LoRA on both ASR encoder and LLM.
% Different from the above simulated data, the training samples in meeting scenarios can have high overlap ratios, which not only lead to high missing error rate in ASR task, but even worse repetition hallucinations for LLM during inference on such data.
% % Since the scarcity of such data, scaling law with huge amount of data is of high cost.
% % , we find that the model probably has when inference on overlapped speech.
% In the next subsection, we investigate the cause and propose a solution to alleviate the hallucinations.

\begin{figure*}[htpb]
    \centering
    \includegraphics[width=0.9\linewidth]{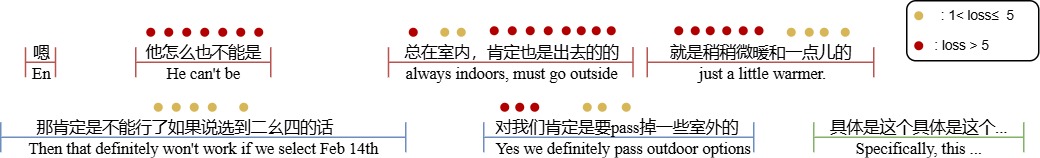}
    \caption{CE loss within overlaps, where the tokens in overlapped regions are of high ASR loss. (Segmented from R003\_M0046 in the Alimeeting training set)}
    \label{fig:hallucinate}
\end{figure*}

\subsection{Hallucination in the overlapped speech}

\begin{CJK}{UTF8}{gbsn}
When training the LLM with overlapped speech, we observe that the trained model tends to generate repeated backchannel words (e.g., ``en'', ``yes'', ``嗯，对'') during inference on overlapped regions. Interestingly, even after removing backchannels from the training transcripts, the hallucination phenomenon persists.

% When using overlapped speech to train LLM,  we find that the model easily generates the repeated backchannel words, such as en, yes (嗯，对) , when inference on the overlaps.
% However, even we remove the backchannels within the overlaps from the training transcripts, the hallucination still exists.
% which motivates us to investigate deeply in the last training stage.
\end{CJK}

We investigate the causes of hallucinations during training.
Figure~\ref{fig:hallucinate} illustrates the ASR loss on a far-field overlapped speech sample in the final training stage. 
Tokens within overlapped regions exhibit significantly higher cross-entropy (CE) loss compared to non-overlapped tokens. 
During training, the model may place disproportionate emphasis on these overlapped tokens while overlooking the relatively easier non-overlapped ones in backpropagation. 
Yet, speech signals from distant speakers in overlapped regions often have very low energy and may be unintelligible even to human listeners. 
In such adverse cases, the LLM can detect the presence of overlaps but cannot reliably infer their content. 
Since overlaps frequently contain backchannel words, the model tends to repeatedly generate these tokens to fill the uncertain regions, thereby triggering hallucinations.

% Figure \ref{fig:hallucinate} illustrates the ASR loss on a far-field overlapped speech sample in the last training stage, where the tokens within the overlaps can have extreme higher CE loss values when compared to the normal loss values on non-overlapped tokens.
% If we directly use these loss for backpropagation,  
% % the losses from the overlapped regions will take the most ratio and make 
% the model may focus on recognizing the overlapped tokens while neglect the easy-recognized non-overlapped tokens.
% However, in far filed recordings, some speech signals from the distant speakers may have very low energy and cannot be recognized accurately  by human when overlapped by the closer speech.
% In such adverse yet frequent cases, the LLM  may only detect the overlaps but cannot guess what it is. 
% Since there are lot of backchannel within overlaps, it prone to generate repetitive backchannel tokens to fill the overlap content, which easily trigger the hallucination.

To address this issue, we propose an adaptive strategy that masks high-loss tokens. 
Specifically, the system automatically select easily recognized tokens for training while discarding overly difficult ones. 
The masking threshold is determined by the distribution of CE loss values within each sample:
% Based on the above conjecture, we carefully mask the high-loss tokens using a adaptive threshold, make the system self select the easy tokens for training and give up the difficult tokens.
% The threshold is determined by the distribution of the CE loss values in the sample, that is,
\begin{equation}
    T_{mask} = max(Avg(\text{CE}(\text{text})), 2.0),
    \label{eq:threshold}
\end{equation}
where the mean of the loss values is used to compute the threshold, and $2.0$ serves as a lower bound. 
During training, loss values exceeding this threshold are masked out. 
The overall training objective is then defined as:
\begin{equation}
    % L = L_{spk} + Mask(L_{asr}, T_{mask})
    L = L_{\text{spk}} + \text{Mask}(L_{\text{asr}}, T_{\text{mask}}),
    \label{eq:all_loss}
\end{equation}
where $\text{Mask}(\cdot)$ excludes the loss values that exceed the threshold $T_{\text{mask}}$.

% Another data preprocess trick is to split the long segments containing pauses into shorter ones according to the the punctuations and then re-alignment the shorter segments based on the force-alignment timestamps, which can better align the labels with the input audio when there exists speaker interrupts.

\section{Experiments}

In the first training stage, we use the WenetSpeech corpus~\cite{zhang2022wenetspeech}. 
% For the second and third stages, we adopt the two-speaker corpus ASR-BIGCCSC\footnote{https://magichub.com/datasets/mandarin-chinese-conversational-speech-corpus-spontaneous-conversation/}, which does not contain overlap labels. 
For the second and third stages, we adopt an internal ASR corpus of approximately 4,000 hours, which consists of long-duration two-speaker conversations with transcriptions but does not contain overlap labels.
Long recordings are truncated into shorter segments (8--20 seconds) according to the reference transcripts, and training samples are simulated by concatenating several segments at most 50 seconds.
In the final stage, we finetune the model on real meeting corpora, AliMeeting~\cite{Yu2022M2MeT} and Aishell4~\cite{AISHELL4}. 
Similar to SpeakerLM~\cite{yin2026speakerlmendtoendversatilespeaker}, only the first channel of the far-field signals is used for both training and evaluation. 
The long audios are split into segments of at most 50 seconds, ensuring that no overlaps occur at the segment boundaries. Table~\ref{tab:data} summarizes the statistics of the training and test speech segments in the final stage,
where the overlap ratio in AliMeeting is substantially higher than that in Aishell4.
In addition, we apply a data preprocessing that splits long segments containing pauses into shorter ones based on punctuation. These shorter segments are then re-aligned using force-alignment timestamps, which improves label alignment with the input audio in cases of speaker interruptions.

% During training, we use WenetSpeech\cite{zhang2022wenetspeech} corpurs in the first stage.
% In the second and third stages, we use the two-speaker corpus ASR-BIGCCSC\footnote{https://magichub.com/datasets/mandarin-chinese-conversational-speech-corpus-spontaneous-conversation} for training, which has no overlap labels. We truncate the long recordings into short segments (8-20 seconds) according to the reference transcripts and simulate training samples by concatenating several segments.
% The real-recorded meeting corpus AliMeeting\cite{Yu2022M2MeT} and Aishell4\cite{AISHELL4} are used in the last stage.
% Similar to SpeakerLM\cite{}, only the first channel of the far-field signals are used for training and evaluation. 
% The long audios are split into at most 50 seconds, and we make sure there is no overlap at two ends of the segments.
% Table \ref{tab:data} shows the statistics of the training and test speech segments in the last stage.
% The overlap ratio in AliMeeting corpus is much higher overlap ratios than that in Aishell4.

% Table generated by Excel2LaTeX from sheet 'Sheet3'
\begin{table}[htbp]
  \centering
  \caption{The statistics of the meeting segments in the last training stage and evaluation stage.}
  \scalebox{0.85}{
    \begin{tabular}{cc|cccc}
    \hline
    \multicolumn{2}{c|}{Source Corpus} & \#Seg. & Avg. Dur.& Avg. \#Spk & Overlap\% \\
    \hline
    {AliMeeting\cite{Yu2022M2MeT}} & Eval  & 297   & 43.82  & 2.96 & 21.42  \\
          & Test  & 737   & 44.78  & 2.85  & 18.80 \\
          (104.75 hours)& Train & 6951  & 43.60  & 3.02 & 25.82  \\
    \hline
    {Aishell4\cite{AISHELL4}} & Eval  & 973   & 45.67  & 2.29 & 5.00 \\
          (107.50 hours)& Train & 7837  & 45.36  & 2.84  & 7.18\\
    \hline
    \hline
    \end{tabular}%
    }
  \label{tab:data}%
\end{table}%

We initialize our model with pretrained modules: SenseVoice-small \cite{FunAudioLLM} for the ASR encoder, Camppus \cite{wang23ha_interspeech} for the speaker feature encoder, and Qwen2.5-0.5B-Instruct~\cite{qwen2.5} as the decoding backend, with a total parameter size of approximately 0.7B. All features are projected into 896 dimensions before being fed into the LLM.E
The learning rates are set to $1\times10^{-4}$ for the first three stages and $2\times10^{-5}$ for the final stage with AdamW optimizer. 
For each stage, the warm-up ratio is fixed at 0.03, and the LoRA rank is set to 16. 
% The parameter $p$ is set to 100\% when computing the masking threshold in Eq.~\eqref{eq:threshold}.
% We use pretrained modules to initialize our model, that is Sensevoice-small for the ASR encoder, Campplus for the speaker feature encoder and Qwen2.5-0.5-Instruct\cite{qwen2.5} for the decoding backend, where the parameter size is around 0.7B.
% The feature dimension are all projected into 512 when input to the LLM.
% The learning rates for four stages are 1e-4 and 2e-5 in the first three training stages and the last stage respectively. 
% For each stage, the warm up ratio is set to 0.03. The LoRA rank is set to 16.
% $p$ is set to to 100\% to compute the mask threshold in Eq\eqref{eq:threshold}.
During inference, we evaluate performance using Character Error Rate (CER), cpCER, and their difference $\Delta$cp. We also set \textit{no\_repeat\_ngram\_size} to 8 to reduce repetition hallucinations.

% During inference, character Error Rate (CER), cpCER and their difference $\Delta$cp are used as evaluation metrics. 
% We set \textit{no\_repeat\_ngram\_size} to 8 to further reduce the repetition hallucination. % to avoid long repetition halluc

For comparison, we construct cascaded pipelines using open-sourced modules: 3D-Speaker~\cite{chen20243d} and Diarizen-large~\cite{han2025efficient} for diarization, and Paraformer-large~\cite{gao22b_interspeech} for ASR. 
We also compare against two end-to-end LLM-based systems: SpeakerLM~\cite{yin2026speakerlmendtoendversatilespeaker} and VibeVoice-ASR~\cite{peng2026vibevoiceasrtechnicalreport}. 
Since SpeakerLM and its test data are not publicly available, we directly report the original results alongside its compared pipeline system. 
For VibeVoice-ASR, we run inference on our evaluation sets and exclude failed samples during inference, where the failed segments account for approximately 2\%.

% For comparison, we use open-sourced modules to compose cascaded pipelines, where  3D-Speaker \cite{chen20243d} and  Diarizen-large \cite{han2025efficient} for diarization and Paraformer-large \cite{gao22b_interspeech} used for ASR.
% We also compare two end-to-end LLM based systems: SpeakerLM \cite{yin2026speakerlmendtoendversatilespeaker} and VibeVoice-ASR\cite{peng2026vibevoiceasrtechnicalreport}. 
% Since SpeakerLM and its test data have not been open-sourced, we directly paste the original results along with its compared pipeline system for comparison.
% For VibeVoice-ASR, we directly use the model to inference on our test sets and remove the failed samples before evaluation, where the failed segments take account around 2\%. %1.3%~2.6%

\section{Results}

% stage3 mix loss			28.27 	37.44 	9.17 	26.62 	37.04 	10.42 	20.56 	28.54 	7.98 

\begin{table*}
    \centering
    \scalebox{0.88}{
% Table generated by Excel2LaTeX from sheet 'Sheet2'
\begin{tabular}{p{12em}c|c|ccc|ccc|ccc}
\hline
\multicolumn{1}{r}{} & \multicolumn{1}{r|}{} & \multicolumn{1}{r|}{ASR} & \multicolumn{3}{c|}{AliMeeting Eval} & \multicolumn{3}{c|}{AliMeeting Test} & \multicolumn{3}{c}{Aishell4 Eval} \\
\multicolumn{1}{l}{System} & \multicolumn{1}{c|}{\#Params} & \multicolumn{1}{c|}{mask} & CER\%   & cpCER\% & $\Delta$cp\% & CER\%   & cpCER\% & $\Delta$cp\%  & CER\%   & cpCER\% & $\Delta$cp\% \\

\hline
Paraformer+3D speaker & \multicolumn{1}{c|}{70M} & \multicolumn{1}{c|}{/} & 31.80  & 36.39  & 4.59  & 27.78  & 32.46  & 4.68  & 22.67  & 28.29  & 5.62 \\
Paraformer+DiariZen-large & 140M  & /     & 31.80  & 36.09  & 4.29  & 27.78  & 33.09  & 5.31  & 22.67  & 26.34  & 3.67 \\
\hdashline
VibeVoice-ASR & 7B    & /     & 31.38  & 39.20  & 7.82  & 29.47  & 35.86  & 6.39  & 21.65  & 26.23  & 4.58 \\
\hdashline
Sensevoice-small & 230M  & /     & 26.62 & /     & /     & 25.08 & /     & /     & 18.86 & /     & / \\
Semantic Feature Only & 0.7B  & $\times$     & 26.66  & 31.11  & 4.45  & 26.12  & 31.94  & 5.82  & 18.38  & 23.08  & 4.70 \\
Feature-wise Concatenation & 0.7B  & $\times$     & 28.30  & 32.23  & 3.93  & 25.60  & 31.47  & 5.87  & 18.73  & 23.54  & 4.81 \\
Temporal Interleave & 0.7B  & $\times$     & 28.07  & 30.77  & 2.70  & 26.22  & 29.71  & \textbf{3.49}  & 18.41  & 21.45  & 3.04 \\
\hdashline
Feature-wise Concatenation & 0.7B  & \checkmark & 26.19  & 29.68  & 3.49  & 24.27  & 29.64  & 5.37  & 17.54  & 22.24  & 4.70 \\
Time-wise Concatenation & 0.7B  & \checkmark     & 25.64  & 28.51  & 2.87  & 25.29  & 30.10  & 4.81  & 17.56  & 20.95  & 3.39 \\
Temporal Interleave & 0.7B  & \checkmark     & 25.56  & \textbf{27.96} & \textbf{2.40} & 23.61  & \textbf{27.16} & 3.55 & 17.18  & \textbf{19.98} & \textbf{2.80} \\
\hline
\multicolumn{1}{l}{\textit{Paraformer+3D speaker* \cite{yin2026speakerlmendtoendversatilespeaker}}} & \multicolumn{1}{c|}{\textit{70M}} & \multicolumn{1}{c|}{/} & /     & /     & /     & \textit{21.30} & \textit{23.20} & \textit{1.90} & \textit{23.02} & \textit{26.01} & \textit{2.99} \\
\multicolumn{1}{l}{\textit{SpeakerLM*   (212 hours) \cite{yin2026speakerlmendtoendversatilespeaker}}} & \multicolumn{1}{c|}{\textit{7B}} & \multicolumn{1}{c|}{/} & /     & /     & /     & \textit{18.63} & \textit{32.22} & \textit{13.59} & \textit{17.75} & \textit{26.14} & \textit{8.39} \\
\multicolumn{1}{l}{\textit{SpeakerLM*   (7638 hours) \cite{yin2026speakerlmendtoendversatilespeaker}}} & \multicolumn{1}{c|}{\textit{7B}} & \multicolumn{1}{c|}{/} & /     & /     & /     & \textit{13.97} & \textit{16.05} & \textit{2.08} & \textit{17.17} & \textit{18.37} & \textit{1.20} \\
\hline
\hline
\multicolumn{12}{l}{
\footnotesize{*: uses a different segment split on the test data from ours.}
% \footnotesize{\+: has different segment split on the test audios from ours.}
} \\

\end{tabular}%
}
    \caption{The performance of systems on test datasets.}
    \label{tab:res}
\end{table*}

Table~\ref{tab:res} presents the performance of different systems across three evaluation sets. The pipeline systems, whether using 3D-Speaker or DiariZen-large for diarization, achieve comparable results. In contrast, the end-to-end model VibeVoice-ASR struggles on highly overlapped segments. 
% likely due to the limited amount of far-field data in its training corpus.

% % Tabel \ref{tab:res} shows the results of different systems on three test sets.
% The pipeline systems using either 3D-speaker or DiaiZen-large as the diarization module have similar performance, while the end-to-end model VibeVoice shows some struggle on the high overlapped segments, which may due to the scare far-field data in the training data (TODO).

Our systems, built upon the SenseVoice-small encoder, inherit its strong ASR capability. 
When using only semantic features as input, the proposed system demonstrates diarization performance comparable to the cascaded baselines. 
Across different feature fusion strategies, we observe that time-wise concatenation and temporal interleave yield better speaker identification accuracy than feature-wise concatenation. 
Because feature-wise concatenation projects semantic and speaker features into a lower-dimensional space, it inevitably reduces the richness of these representations and compromises their completeness. 
By contrast, temporal merging approaches preserve speaker information more effectively. 
In particular, the temporal interleave format, which relies on positional encoding between adjacent segments, aligns more naturally with our label structure and achieves the best performance on $\Delta$cp.

% In particular, the temporal interleave format, which relies solely on positional encoding between adjacent segments, enhances recognition of both transcripts and speaker identities within the same segment, achieving the best performance on $\Delta$cp.

% Our systems composed of the SenseVoice-small encoder inherit the original ASR capability.
% When using only semantic features as input, the proposed system can shows comparable diarization capability with the cascaded baseline system.
% Across the different feature merge strategies, we find that the time-wise concatenation and temporal interleave approaches show better speaker identification capability than that in the feature-wise concatenation approach.
% Although the feature-wise concatenation can save the input length, it may undermined the completeness of the semantic features and speaker features by projecting them into a lower-dimension space.
% The speaker features in temporal merge approaches shows its effectiveness in speaker ID identification.
% The temporal interleave format using only positional encoding can better benefit the recognition of both the transcripts and the corresponding speaker ID for each segment and shows the best performance on $\Delta$cp.
% cope with the labels in Eq \eqref{eq:label}.
% Moreover, we compare the another dual-encoder without interleaving, but insert digital anchors, our methods is still better.
% In the next subsection, We also investigate the performance when reducing input lengths.

Applying the adaptive mask threshold defined in Eq.~\eqref{eq:threshold} further alleviates repetition hallucinations in contextual recognition. 
This improves both ASR and diarization performance, with relative cpCER gains of 8.5\% on the AliMeeting test set and 6.9\% on the Aishell4 evaluation set.

% With the ASR threshold in Eq\eqref{eq:threshold} applied on the loss function, the repetition hallucination in context recognition can be greatly alleviated, where the ASR capability and diarization capability can be further imported and the relatively improvements of cpCER can achieve 8.5\% on the AliMeeting test set and 6.9\% on the Aishell4 evaluation set.

Since our evaluation sets differ from those used in SpeakerLM~\cite{yin2026speakerlmendtoendversatilespeaker}, we also report their results alongside the same baseline system (\textit{Paraformer+3D speaker}). 
This comparison indicates that our evaluation sets are more challenging. 
Nevertheless, our system outperforms SpeakerLM trained on the same 212 hours dataset and achieves comparable $\Delta$cp results to SpeakerLM trained with 7,638 hours of meeting data.

% Since Our test sets are different from those in the SpeakerLM\cite{yin2026speakerlmendtoendversatilespeaker}, we paste their results using the same baseline system (\textit{Paraformer+3D speaker}), which shows that our evaluation sets are more difficult.
% In such cases, our system still outperforms the SpeakerLM using the same training dataset (212 hours) and get comparable results on $\Delta$cp to the SpeakerLM trained with 7638 hours of meeting data.

\subsection{Ablation study on the length of the speaker feature}

\begin{figure}
    \centering
    \includegraphics[width=1.0\linewidth]{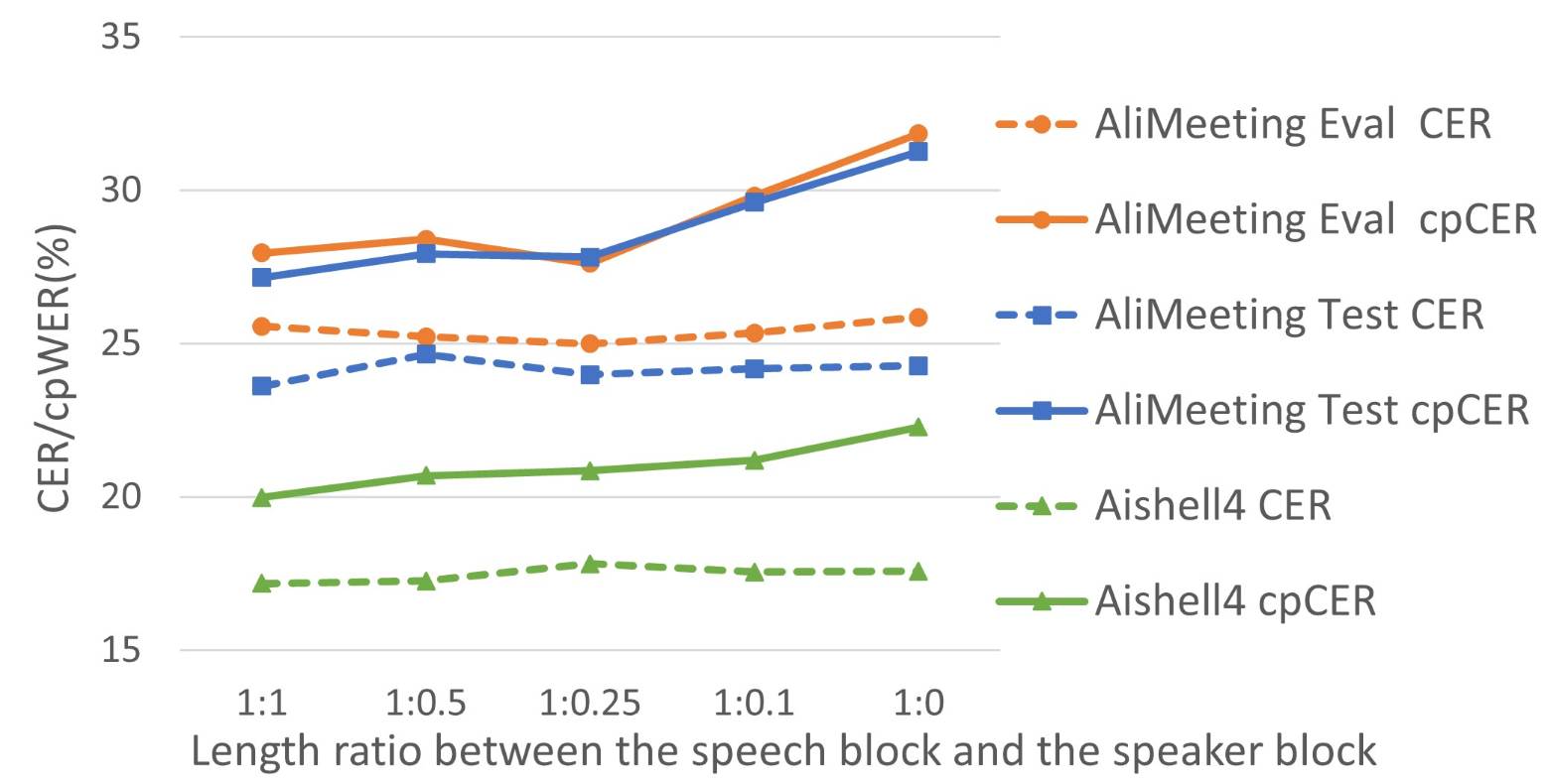}
    % \caption{The performance of system with different down sampling rate on speaker feature seqneces.}
    \caption{System performance with different downsampling rates applied to speaker feature sequences.}
    \label{fig:downsample}
\end{figure}

Since the temporal interleave strategy doubles the input length during inference, we explore downsampling the frames in each speaker chunk during the final training stage to reduce sequence length and computational cost.
As shown in Figure~\ref{fig:downsample}, shortening the speaker feature sequence does not affect ASR capability but degrades speaker identification, where CER remains stable while cpCER increases. 
When downsampling the speaker feature sequence by a factor of four (i.e., ratio of 1:0.25), cpCER remains within an acceptable range while reducing the input length by 75\%. 
Even with only 10\% of the speaker frames retained, the system still provides significant improvements compared to using semantic features alone.
This suggests that moderate downsampling offers a good trade-off between efficiency and performance.

We also attempted frame-level interleaving (N=1), but the training loss failed to converge.

% Since the temporal interleave strategy doubles the input lengths during inference, to reduce the input cost, we simply down sample the frames in each speaker block in the last training stage.
% As shown in Figure \ref{fig:downsample}, we find that shortening the speaker features will not affect the ASR capability but affect the speaker identification, where the CER keeps stable and cpCER increases. 
% When down sampling the speaker feature sequence of 4 times (1:0.25), the cpCER results are still acceptable which can save 75\% input length.
% Even with 10\% speaker frames, it can still provide significant improvement compared with the system using only semantic feature.

\subsection{Ablation study on the loss function}

\begin{table}
    \centering
% Table generated by Excel2LaTeX from sheet 'Sheet2'
\scalebox{0.82}{
\begin{tabular}{l|ccc}
\hline
\multicolumn{1}{r|}{} & AliMeet. Eval& AliMeet. Test& Aishell4 Eval \\
\hline
Temporal Interleave & 25.56/\textbf{27.96} & 23.61/\textbf{27.16} & 17.18/\textbf{19.98} \\
\multicolumn{1}{l|}{w.o. Speaker loss} & 25.20/28.76 & 23.80/28.54 & 16.95/20.05 \\
w.o. ASR mask (T=$\infty$)& 28.07/30.77 & 26.22/29.71 & 18.41/21.45 \\
w.o. ASR loss (T=0) & 28.57/31.31 & 26.59/30.34 & 28.94/33.17 \\
w.o. seg. realignment & 25.67/28.74 & 24.77/28.95 & 17.32/20.26 \\
\hline
\hline
\end{tabular}%
}
    \caption{Ablation study by removing the proposed strategies (CER/cpCER\%). $T$ is the threshold defined in Eq.~\eqref{eq:threshold}.}
    \label{tab:remove}
\end{table}

To demonstrate the effectiveness of our proposed loss function in Eq.~\eqref{eq:spkloss}, we train a system using only the ASR loss, with results shown in the second row of Table~\ref{tab:remove}. 
We also evaluate different masking thresholds $T$ in Eq.~\eqref{eq:threshold} and Eq.~\eqref{eq:all_loss}. 
When $T=\infty$, no masking is applied and the original ASR loss is used in the final training stage. 
When $T=0$, all ASR loss values are masked and only the $L_{spk}$ is used. 
Neither yields satisfactory results, whereas setting a reasonable threshold provides a better balance. 
Finally, the last row of Table~\ref{tab:remove} shows that segment re-alignment of transcripts in data preprocessing also benefits training, which better matchs the input feature sequences and the target label sequences.

% To show the effectiveness of our proposed loss function in Eq\eqref{eq:spkloss}, we train a system only using ASR loss and the results are shown in the second row in Table \ref{tab:remove}.
% We also show the results using different masking thresholds $T$ in Eq\ref{eq:threshold} and Eq\ref{eq:all_loss} in Table \ref{tab:remove}.
% When the threshold equals to infinite, no mask is applied and original ASR loss is used in the last training stage.
% When the threshold becomes zero, all ASR loss values are masked and only speaker ID loss is used in  training.
% These two opposite approaches do not have idea results, while give a reasonable threshold can be a better choice.
% The last row in the Table \ref{tab:remove} shows that the segment re-alignment of transcripts also benefit the training, which makes  the input and the output match to each other along the temporal axis.

\section{Conclusions}

In this paper, we proposed an end-to-end model for multi-speaker ASR with only 0.7B parameters, which achieves strong performance on far-field recordings with high overlap ratios. 
The relationship between ASR and diarization tasks is carefully balanced through both the model architecture and the training scheme. 
Our experiments demonstrate that adaptive loss masking and temporal interleave strategies effectively mitigate hallucinations and improve speaker attribution. 
Future work will explore speaker registration and longer timestamp modeling to enhance multi-talker dialogue understanding.

% We proposed a compact end-to-end model for multi-speaker ASR with 0.7B parameters, achieving strong performance on far-field recordings with high overlap ratios. By balancing ASR and diarization tasks through architecture and training design, the system effectively mitigates hallucinations and improves speaker attribution. 

% In this paper, we proposed an end-to-end model for multi-speaker ASR task with only 0.7B parameters, which shows excellent performance on far-field recordings of high overlap ratios.
% The relationship between ASR task and diarization task are well balanced during both model structure and training schema.
% In the future work, we consider to extend the model capability with registering the speakers in advance which more cope with multi-talker dialogue understanding. 
% Longer input and output timestamps capabilities will also be considered. 

\section{Generative AI Use Disclosure}
During the preparation of this manuscript, generative AI tools were used solely for language editing and manuscript polishing to enhance readability. These tools were not employed to generate scientific ideas, experimental data, or technical contributions. All authors have thoroughly reviewed and approved the final version of the manuscript and assume full responsibility for the integrity and completeness of its content.

% \section{Acknowledgments}

% {\color{blue}Acknowledgments should be included only in the camera-ready version, not in the version submitted for review. For regular papers, pages 5 and 6, and for long papers, pages 9 and 10, are reserved exclusively for acknowledgments, disclosures of the use of generative AI tools, and references. No other content may appear on these pages. Any appendices must be contained within the first four pages for regular papers and within the first eight pages for long papers.

% Acknowledgments and references may begin on an earlier page if space permits.}

% \ifcameraready
%      The Interspeech 2026 organizers
% \else
%      The authors
% \fi
% would like to thank ISCA and the organizing committees of past Interspeech conferences for their help and for kindly providing the previous version of this template.

\bibliographystyle{IEEEtran}
\bibliography{mybib}

@inproceedings{wang24h_interspeech,
  title     = {{DiarizationLM: Speaker Diarization Post-Processing with Large Language Models}},
  author    = {Quan Wang and Yiling Huang and Guanlong Zhao and Evan Clark and Wei Xia and Hank Liao},
  year      = {2024},
  booktitle = {{Interspeech 2024}},
  pages     = {3754--3758},
  doi       = {10.21437/Interspeech.2024-209},
  issn      = {2958-1796},
}

@misc{lin2025diarizationawaremultispeakerautomaticspeech,
      title={Diarization-Aware Multi-Speaker Automatic Speech Recognition via Large Language Models}, 
      author={Yuke Lin and Ming Cheng and Ze Li and Beilong Tang and Ming Li},
      year={2025},
      eprint={2506.05796},
      archivePrefix={arXiv},
      primaryClass={eess.AS}, 
}

@misc{shi2025trainshortinferlong,
      title={Train Short, Infer Long: Speech-LLM Enables Zero-Shot Streamable Joint ASR and Diarization on Long Audio}, 
      author={Mohan Shi and Xiong Xiao and Ruchao Fan and Shaoshi Ling and Jinyu Li},
      year={2025},
      eprint={2511.16046},
      archivePrefix={arXiv},
      primaryClass={eess.AS},
      url={https://arxiv.org/abs/2511.16046}, 
}

@misc{yin2026speakerlmendtoendversatilespeaker,
      title={Speaker{LM}: End-to-End Versatile Speaker Diarization and Recognition with Multimodal Large Language Models}, 
      author={Han Yin and Yafeng Chen and Chong Deng and Luyao Cheng and Hui Wang and Chao-Hong Tan and Qian Chen and Wen Wang and Xiangang Li},
      year={2026},
      eprint={2508.06372},
      archivePrefix={arXiv},
      primaryClass={cs.SD},
      url={https://arxiv.org/abs/2508.06372}, 
}

@misc{ai2026mosstranscribediarizetechnical,
      title={{MOSS} Transcribe Diarize Technical Report}, 
      author={MOSI. AI and Donghua Yu and Zhengyuan Lin and Chen Yang and Yiyang Zhang and Hanfu Chen and Jingqi Chen and Ke Chen and Liwei Fan and Yi Jiang and Jie Zhu and Muchen Li and Wenxuan Wang and Yang Wang and Zhe Xu and Yitian Gong and Yuqian Zhang and Wenbo Zhang and Songlin Wang and Zhiyu Wu and Zhaoye Fei and Qinyuan Cheng and Shimin Li and Xipeng Qiu},
      year={2026},
      eprint={2601.01554},
      archivePrefix={arXiv},
      primaryClass={cs.SD},
      url={https://arxiv.org/abs/2601.01554}, 
}

@INPROCEEDINGS{9383556,
  author={Raj, Desh and Denisov, Pavel and Chen, Zhuo and Erdogan, Hakan and Huang, Zili and He, Maokui and Watanabe, Shinji and Du, Jun and Yoshioka, Takuya and Luo, Yi and Kanda, Naoyuki and Li, Jinyu and Wisdom, Scott and Hershey, John R.},
  booktitle={2021 IEEE Spoken Language Technology Workshop (SLT)}, 
  title={Integration of Speech Separation, Diarization, and Recognition for Multi-Speaker Meetings: System Description, Comparison, and Analysis}, 
  year={2021},
  volume={},
  number={},
  pages={897-904},
  doi={10.1109/SLT48900.2021.9383556}}

@INPROCEEDINGS{7953097,
  author={Hruz, Marek and Zajic, Zbynek},
  booktitle={2017 IEEE International Conference on Acoustics, Speech and Signal Processing (ICASSP)}, 
  title={Convolutional Neural Network for speaker change detection in telephone speaker diarization system}, 
  year={2017},
  volume={},
  number={},
  pages={4945-4949},
  doi={10.1109/ICASSP.2017.7953097}}

@INPROCEEDINGS{10037902,
  author={Lin, Yuxiao and Du, Zhihao and Zhang, Shiliang and Yu, Fan and Zhao, Zhou and Wu, Fei},
  booktitle={2022 13th International Symposium on Chinese Spoken Language Processing (ISCSLP)}, 
  title={Separate-to-Recognize: Joint Multi-target Speech Separation and Speech Recognition for Speaker-attributed ASR}, 
  year={2022},
  volume={},
  number={},
  pages={150-154},
  doi={10.1109/ISCSLP57327.2022.10037902}}

@ARTICLE{7122291,
  author={Weng, Chao and Yu, Dong and Seltzer, Michael L. and Droppo, Jasha},
  journal={IEEE/ACM Transactions on Audio, Speech, and Language Processing}, 
  title={Deep Neural Networks for Single-Channel Multi-Talker Speech Recognition}, 
  year={2015},
  volume={23},
  number={10},
  pages={1670-1679},
  doi={10.1109/TASLP.2015.2444659}}

@ARTICLE{9072433,
  author={Zhang, Wangyou and Chang, Xuankai and Qian, Yanmin and Watanabe, Shinji},
  journal={IEEE/ACM Transactions on Audio, Speech, and Language Processing}, 
  title={Improving End-to-End Single-Channel Multi-Talker Speech Recognition}, 
  year={2020},
  volume={28},
  number={},
  pages={1385-1394},
  doi={10.1109/TASLP.2020.2988423}}

@misc{huo2026tagspeechendtoendmultispeakerasr,
      title={{TagSpeech}: End-to-End Multi-Speaker ASR and Diarization with Fine-Grained Temporal Grounding}, 
      author={Mingyue Huo and Yiwen Shao and Yuheng Zhang},
      year={2026},
      eprint={2601.06896},
      archivePrefix={arXiv},
      primaryClass={eess.AS},
      url={https://arxiv.org/abs/2601.06896}, 
}

@misc{peng2026vibevoiceasrtechnicalreport,
      title={{VibeVoice-ASR} Technical Report}, 
      author={Zhiliang Peng and Jianwei Yu and Yaoyao Chang and Zilong Wang and Li Dong and Yingbo Hao and Yujie Tu and Chenyu Yang and Wenhui Wang and Songchen Xu and Yutao Sun and Hangbo Bao and Weijiang Xu and Yi Zhu and Zehua Wang and Ting Song and Yan Xia and Zewen Chi and Shaohan Huang and Liang Wang and Chuang Ding and Shuai Wang and Xie Chen and Furu Wei},
      year={2026},
      eprint={2601.18184},
      archivePrefix={arXiv},
      primaryClass={cs.SD},
      url={https://arxiv.org/abs/2601.18184}, 
}

@INPROCEEDINGS{METACATwang,
  author={Wang, Jinhan and Wang, Weiqing and Dhawan, Kunal and Park, Taejin and Kim, Myungjong and Medennikov, Ivan and Huang, He and Koluguri, Nithin and Balam, Jagadeesh and Ginsburg, Boris},
  booktitle={ICASSP 2025 - 2025 IEEE International Conference on Acoustics, Speech and Signal Processing (ICASSP)}, 
  title={{META-CAT}: Speaker-Informed Speech Embeddings via Meta Information Concatenation for Multi-talker ASR}, 
  year={2025},
  volume={},
  number={},
  pages={1-5},
  doi={10.1109/ICASSP49660.2025.10889841}}

@InProceedings{Sortformer,
  title = 	 {Sortformer: A Novel Approach for Permutation-Resolved Speaker Supervision in Speech-to-Text Systems},
  author =       {Park, Taejin and Medennikov, Ivan and Dhawan, Kunal and Wang, Weiqing and Huang, He and Koluguri, Nithin Rao and Puvvada, Krishna C and Balam, Jagadeesh and Ginsburg, Boris},
  booktitle = 	 {Proceedings of the 42nd International Conference on Machine Learning},
  pages = 	 {48153--48169},
  year = 	 {2025},
  editor = 	 {Singh, Aarti and Fazel, Maryam and Hsu, Daniel and Lacoste-Julien, Simon and Berkenkamp, Felix and Maharaj, Tegan and Wagstaff, Kiri and Zhu, Jerry},
  volume = 	 {267},
  series = 	 {Proceedings of Machine Learning Research},
  month = 	 {13--19 Jul},
  publisher =    {PMLR}
}

@article{FunAudioLLM,
  publtype={informal},
  author={Keyu An and Qian Chen and Chong Deng and Zhihao Du and Changfeng Gao and Zhifu Gao and Yue Gu and Ting He and Hangrui Hu and Kai Hu and Shengpeng Ji and Yabin Li and Zerui Li and Heng Lu and Haoneng Luo and Xiang Lv and Bin Ma and Ziyang Ma and Chongjia Ni and Changhe Song and Jiaqi Shi and Xian Shi and Hao Wang and Wen Wang and Yuxuan Wang and Zhangyu Xiao and Zhijie Yan and Yexin Yang and Bin Zhang and Qinglin Zhang and Shiliang Zhang and Nan Zhao and Siqi Zheng},
  title={{FunAudioLLM}: Voice Understanding and Generation Foundation Models for Natural Interaction Between Humans and LLMs},
  year={2024},
  cdate={1704067200000},
  journal={CoRR},
  volume={abs/2407.04051},
}

@inproceedings{wang23ha_interspeech,
  title     = {{CAM++: A Fast and Efficient Network for Speaker Verification Using Context-Aware Masking}},
  author    = {Hui Wang and Siqi Zheng and Yafeng Chen and Luyao Cheng and Qian Chen},
  year      = {2023},
  booktitle = {{Interspeech 2023}},
  pages     = {5301--5305},
  doi       = {10.21437/Interspeech.2023-1513},
  issn      = {2958-1796},
}

@inproceedings{hu2022lora,
title={Lo{RA}: Low-Rank Adaptation of Large Language Models},
author={Edward J Hu and Yelong Shen and Phillip Wallis and Zeyuan Allen-Zhu and Yuanzhi Li and Shean Wang and Lu Wang and Weizhu Chen},
booktitle={International Conference on Learning Representations},
year={2022}
}

@inproceedings{zhang2022wenetspeech,
  title={{WenetSpeech}: A 10000+ Hours Multi-domain Mandarin Corpus for Speech Recognition},
  author={Zhang, Binbin and Lv, Hang and Guo, Pengcheng and Shao, Qijie and Yang, Chao and Xie, Lei and Xu, Xin and Bu, Hui and Chen, Xiaoyu and Zeng, Chenchen and Wu, Di and Peng, Zhendong},
  booktitle={International Conference on Acoustics, Speech and Signal Processing (ICASSP)},
  year={2022},
  organization={IEEE}
}

@inproceedings{Yu2022M2MeT,
  title={M2{M}e{T}: The {ICASSP} 2022 Multi-Channel Multi-Party Meeting Transcription Challenge},
  author={Yu, Fan and Zhang, Shiliang and Fu, Yihui and Xie, Lei and Zheng, Siqi and Du, Zhihao and Huang, Weilong and Guo, Pengcheng and Yan, Zhijie and Ma, Bin and Xu, Xin and Bu, Hui},
  booktitle={Proc. ICASSP},
  year={2022},
  organization={IEEE}
}

@inproceedings{AISHELL4,
title={{AISHELL-4}: An Open Source Dataset for Speech Enhancement, Separation, Recognition and Speaker Diarization in Conference Scenario},
author={Yihui, Fu and Luyao, Cheng and Shubo, Lv and Yukai, Jv and Yuxiang, Kong and Zhuo, Chen and Yanxin, Hu and Lei, Xie and Jian, Wu and Hui, Bu and Xin, Xu and Jun, Du and Jingdong, Chen},
booktitle={Interspeech},
year={2021}
}

@misc{qwen2.5,
    title = {Qwen2.5: A Party of Foundation Models},
    url = {https://qwenlm.github.io/blog/qwen2.5},
    author = {Qwen Team},
    month = {September},
    year = {2024}
}

@article{chen20243d,
  title={{3D-Speaker-Toolkit}: An Open Source Toolkit for Multi-modal Speaker Verification and Diarization},
  author={Chen, Yafeng and Zheng, Siqi and Wang, Hui and Cheng, Luyao and others},
  booktitle={ICASSP},
  year={2025}
}

@article{han2025efficient,
  title={Efficient and Generalizable Speaker Diarization via Structured Pruning of Self-Supervised Models},
  author={Han, Jiangyu and P{\'a}lka, Petr and Delcroix, Marc and Landini, Federico and Rohdin, Johan and Cernock{\`y}, Jan and Burget, Luk{\'a}{\v{s}}},
  journal={arXiv preprint arXiv:2506.18623},
  year={2025}
}

@inproceedings{gao22b_interspeech,
  author={Zhifu Gao and ShiLiang Zhang and Ian McLoughlin and Zhijie Yan},
  title={Paraformer: Fast and Accurate Parallel Transformer for Non-autoregressive End-to-End Speech Recognition},
  year=2022,
  booktitle={Proc. Interspeech 2022},
  pages={2063--2067}
}

@INPROCEEDINGS{SCDiar,
  author={Zheng, Naijun and Wan, Xucheng and Liu, Kai and Huan, Zhou},
  booktitle={ICASSP 2025 - 2025 IEEE International Conference on Acoustics, Speech and Signal Processing (ICASSP)}, 
  title={{SCDiar}: a streaming diarization system based on speaker change detection and speech recognition}, 
  year={2025},
  volume={},
  number={},
  pages={1-5}
  }

@article{Watanabe2020CHiME6CT,
  title={{CHiME}-6 Challenge: Tackling Multispeaker Speech Recognition for Unsegmented Recordings},
  author={Shinji Watanabe and Michael Mandel and Jon Barker and Emmanuel Vincent},
  journal={ArXiv},
  year={2020},
  volume={abs/2004.09249}
}

@article{Kanda2021TranscribetoDiarizeNS,
  title={Transcribe-to-Diarize: Neural Speaker Diarization for Unlimited Number of Speakers Using End-to-End Speaker-Attributed ASR},
  author={Naoyuki Kanda and Xiong Xiao and Yashesh Gaur and Xiaofei Wang and Zhong Meng and Zhuo Chen and Takuya Yoshioka},
  journal={ICASSP 2022 - 2022 IEEE International Conference on Acoustics, Speech and Signal Processing (ICASSP)},
  year={2021},
  pages={8082-8086},
  url={https://api.semanticscholar.org/CorpusID:238419307}
}
% \end{CJK}

\end{document}